\documentclass{siamltex}
\usepackage{euscript,amsmath,amssymb,amsfonts,graphicx,bm}
  \usepackage{epstopdf}
  \usepackage{graphics} 
 \usepackage[colorlinks=true]{hyperref}
 \hypersetup{urlcolor=blue, citecolor=red}
 \usepackage[latin1]{inputenc}
\newcommand{\calU}{{\mathcal U}}
\newcommand{\R}{{\mathbb R}}

\newcommand{\X}{\mathbf{X}}

\newcommand{\x}{\mathbf{x}}

\newcommand{\y}{\mathbf{y}}
\newcommand{\e}{{\mathrm e}}
\newcommand{\E}{{\mathbb E}}

\newcommand{\n}{\mathbf n}
\newcommand{\calT}{{\mathcal T}}

\newcommand{\p}{\widetilde{p}}
\newcommand{\q}{{\widetilde{q}}}
\newcommand{\Markov}[2]{\underset{#1}{\overset{#2}{\rightleftharpoons}}}

\begin{document}

\title{The narrow capture problem with partially absorbing targets and stochastic resetting}

\author{Paul C. Bressloff\thanks{Department of Mathematics, University of Utah, Salt Lake City, UT 84112
USA ({\tt bressloff@math.utah.edu})} \and Ryan D. Schumm\thanks{Department of Mathematics, University of Utah, Salt Lake City, UT 84112
USA} }

\date{today}

 \maketitle

\begin{abstract} 
We consider a particle undergoing diffusion with stochastic resetting in a bounded domain $\calU\subset \R^d$ for $d=2,3$. The domain is perforated by a set of partially absorbing targets within which the particle may be absorbed at a rate $\kappa$. Each target is assumed to be much smaller than $|\calU|$, which allows us to use asymptotic and Green's function methods to solve the diffusion equation in Laplace space. In particular, we construct an inner solution within the interior and local exterior of each target, and match it with an outer solution in the bulk of $\calU$. This yields an asymptotic expansion of the Laplace transformed flux into each target in powers of $\nu=-1/\ln \epsilon$ ($d=2$) and $\epsilon$ ($d=3$),  respectively, where $\epsilon$ is the non-dimensionalized target size. The fluxes determine how the mean first-passage time to absorption depends on the reaction rate $\kappa$ and the resetting rate $r$. For a range of parameter values, the MFPT is a unimodal function of $r$, with a minimum at an optimal resetting rate $r_{\rm opt}$ that depends on $\kappa$ and the target configuration.  
\end{abstract}

\section{Introduction}
\begin{figure}[t!]
  \centering
  \includegraphics[width=10cm]{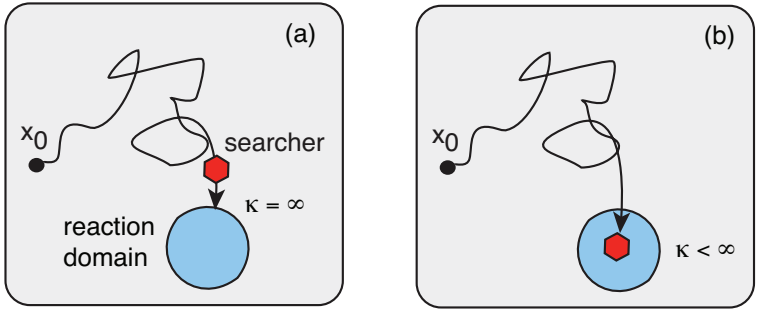}
  \caption{Two models of a chemical reaction. (a) A diffusing particle reacts as soon as it reaches the boundary of the reaction domain. (b) A particle can diffuse in and out of the reaction domain and reacts at a finite rate $\kappa$ within the domain. In the presence of stochastic resetting, the particle can instantaneously return to its initial point $\x_0$ at a rate $r$ prior to reacting, after which the search process is restarted.}
  \label{fig1}
\end{figure}

Chemical reaction kinetics depend on the transport mechanisms that enable two reactants A and B to meet within their reaction radius \cite{Rice85}. The transport process can be formulated as a searcher A looking for a target B, and the effective reaction rate related to a first passage time (FPT) problem \cite{Hanggi90}. Within the interior of living cells, molecular concentrations tend to be dilute so that there are very few copies of reactants. This means that the corresponding search process involves finding a small hidden target within a much larger bounded domain -- the so-called narrow capture problem. The small size of the targets can be exploited to solve the FPT problem using matched asymptotic expansions and Green's function methods \cite{Ward93,Bressloff08,Coombs09,Cheviakov11,Chevalier11,Coombs15,Ward15,Lindsay15,Lindsay17,Grebenkov20}.  Given the fact that passive diffusion results in unrealistically slow reaction rates, there has been considerable interest in identifying both natural and artificial mechanisms for speeding up the underlying search process. One major example is random intermittent transport, in which a reactant switches between passive diffusion and active ballistic motion \cite{Loverdo08,Benichou10}; the latter could be mediated by molecular motors transiently binding diffusing reactant molecules within the cellular environment. One finds that intermittent transport can significantly enhance chemical reactivity and often leads to a non-trivial optimization of reaction rates. (The general theory of random intermittent search processes is reviewed in \cite{Benichou11}.) An idealized version of a random intermittent search process is diffusion with stochastic resetting or restart, whereby the position of a Brownian particle is reset to a fixed location at a random sequence of times, which is typically (but not necessarily) generated by a Poisson process \cite{Evans11a,Evans11b,Evans14}. Moreover, there typically exists an optimal resetting rate for minimizing the mean first passage time (MFPT) to find a target. The simplicity of the resetting protocol means that it can be applied to a wide range of stochastic processes beyond Brownian motion, see the recent review \cite{Evans20} and references therein. There are a number of different reaction scenarios that can be combined with the transport process, as illustrated in Fig. \ref{fig1}. In the case of a diffusion-limited reaction, as soon as the particle hits the boundary of the reaction domain or target it reacts instantaneously (totally absorbing target boundary); alternatively, there could be a nonzero probability that the particle is reflected rather than absorbed (partially absorbing target boundary). These are the typical scenario assumed in narrow-capture problems. However, another possibility is that the particle diffuses in and out of the reaction domain, and reacts at a finite rate $\kappa$ when inside the domain. The target thus acts like a partially absorbing chemical substrate.

Previous studies of the narrow capture problem have focused primarily on partially or totally absorbing interior boundaries without stochastic resetting. However, we have recently shown how asymptotic methods can be extended to incorporate the effects of stochastic resetting by solving the diffusion equation in Laplace space and determining the resulting flux into each target \cite{Bressloff21A,Bressloff21B}; the Laplace transformed flux acts as a generator of statistical quantities such as the MFPT to absorption. In this paper, we analyze the two-dimensional (2D) and three-dimensional (3D) narrow capture problem with stochastic resetting and partially absorbing targets. We proceed by considering the diffusion equation in a bounded, perforated domain $\calU\subset \R^d$, $d=2,3$. Working in Laplace space, we construct an inner solution that holds within the interior and local exterior of each reaction domain, and then match it with an outer solution in the bulk $\calU$. This yields an asymptotic expansion of the Laplace transformed flux into each reaction domain in powers of $\nu =-1/\ln \epsilon$ ($d=2$) and $\epsilon$ ($d=3$), where $\epsilon$ is the non-dimensionalized target size. The Laplace transformed fluxes are used to determine how the MFPT to absorption depends on the reaction rate $\kappa$ and the resetting rate $r$. For a range of parameter values, we find that the MFPT is a unimodal function of $r$, with a minimum at an optimal resetting rate $r_{\rm opt}$ that depends on $\kappa$ and the target configuration. Finally, we consider a more general chemical reaction scheme by taking each target to be a chemical substrate to which a particle can reversibly bind and sequence through a set of intermediate states before being irreversibly absorbed. The narrow escape problem is formulated in \S 2, the matched asymptotic expansions for $d=2$ and $d=3$ are carried out in \S 3 and \S 4, respectively.

\section{The narrow capture problem}
\begin{figure}[b!]
\centering
\includegraphics[width=12cm]{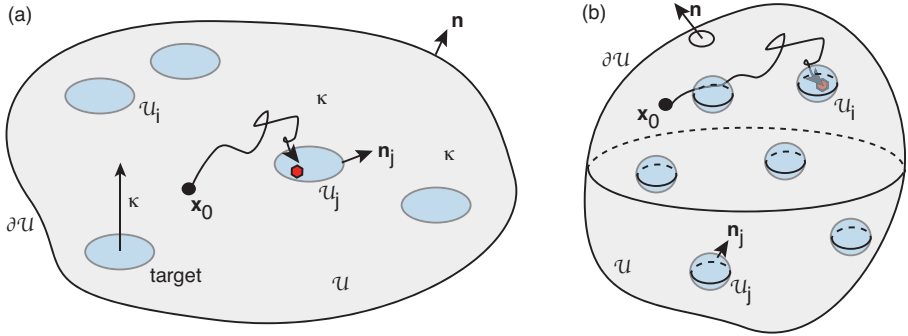} 
\caption{Random search in a domain $\calU\subseteq \R^d$ with $N$ partially absorbing targets $\calU_j$, $j=1,\ldots,N$. Whenever the searcher is within the target domain $\calU_j$, it can be absorbed at a rate $\kappa$. (a) $d=2$, (b) $d=3$.}
\label{fig2}
\end{figure}

Consider a set of partial absorbing targets $\calU_k\subset \calU$, $k=1,\ldots,N$, in a bounded search domain $\calU\subset \R^d$ and set $\bigcup_{k=1}^N \calU_k=\calU_a$, see Fig. \ref{fig2}. Whenever the particle is within $\calU_k$ it can be absorbed (react) at a rate $\kappa$. Each target is taken to be much smaller than $\calU$, that is, $|\calU_j|\sim \epsilon^d |\calU|$ with $\calU_j\rightarrow \x_j\in \calU$ uniformly as $\epsilon \rightarrow 0$, $j=1,\ldots,N$. In addition, the targets are assumed to be well separated with $|\x_i-\x_j|=O(1)$, $j\neq i$, and $\mbox{dist}(x_j,\partial \calU)=O(1)$ for all $i=1,\ldots,N$. In order to develop the analysis we will assume, for simplicity, that each target is a $d$-dimensional sphere of radius $\epsilon \ell_j$: $\calU_i=\{\x \in \calU, \ |\x-\x_i|\leq \epsilon \ell_i\}$.

Let $p(\x,t|\x_0)$ be the probability density that at time $t$ a particle is at $\X(t)=\x$, having started at position $\x_0$. We will set $p=q$ for all $\x\in \calU\backslash \calU_a$ and $p=p_k$ for all $\x\in \calU_k$ such that 
\begin{subequations} 
\label{master}
\begin{align}
	\frac{\partial q(\x,t|\x_0)}{\partial t} &= D\nabla^2 q(\x,t|\x_0), \ \x\in \calU \backslash \calU_a,\\
	\frac{\partial p_k(\x,t|\x_0)}{\partial t} &= D\nabla^2 p_k(\x,t|\x_0) -\kappa p_k(\x,t|\x_0),\ \x\in \calU_k,
	\end{align}
	\end{subequations}
together with continuity conditions at each target boundary,
\begin{equation}
\label{masterc}
 q(\x,t|\x_0)=p_k(\x,t|\x_0),\ \nabla q(\x,t|\x_0)\cdot \n_k =  \nabla p_k(\x,t|\x_0)\cdot \n_k
\end{equation}
for all $\x\in \partial \calU_k$,
and the exterior boundary condition
\begin{equation}
\nabla q\cdot \n=0,\ \x \in \partial \calU .
\end{equation}
Here $\kappa$ is the rate at which the particle is absorbed by a target, $\n$ is the outward unit normal at a point on $\partial\calU$, and $\n_k$ is the outward unit normal at a point on $\calU_i$. We will assume that the particle starts outside all of the targets, so that
\begin{equation}
q(\x,0|\x_0)=\delta(\x-\x_0),\quad p_k(\x,0|\x_0)=0.
\end{equation}
It follows that in the limit $\kappa \rightarrow \infty$, the particle is immediately absorbed when it hits any target boundary $\partial \calU_i$. The latter then acts as a perfect absorber.

The probability flux into the $k$-th target at time $t$ is 
\begin{align}
\label{J}
	J_k(\x_0,t)&= \kappa\int_{\calU_k} p_k(\x,t|\x_0)d\x,\ k = 1,\ldots,N.
	\end{align}
Hence, the splitting probability that the particle is eventually captured by the $k$-th target is
\begin{equation}
\label{split}
\pi_k(\x_0)=\int_0^{\infty}J_k(\x_0,t')dt' =\widetilde{J}_k(\x_0,0),
\end{equation}
where $\widetilde{J}_k(\x_0,s)$ denotes the Laplace transform of $J_k(\x_0,t)$.
Let $Q(\x_0,t)$ denote the survival probability that the particle hasn't been absorbed by a target in the time interval $[0,t]$, having started at $\x_0$:
\begin{align}
\label{Q1}
 Q(\x_0,t)=\int_{\calU}p(\x,t|\x_0)d\x&=\int_{\calU\backslash \calU_a}q(\x,t|\x_0)d\x+\sum_{k=1}^N\int_{\calU_k}p_k(\x,t|\x_0)d\x. \end{align}
Differentiating both sides of this equation with respect to $t$ and using Eqs. (\ref{master}a,b) implies that
\begin{align}
 \frac{\partial Q(\x_0,t)}{\partial t}&=D\int_{\calU\backslash \calU_a}\nabla^2q(\x,t|\x_0)d\x +\sum_{k=1}^N\int_{\calU_k}\left [D\nabla^2 p_k(\x,t|\x_0)-\kappa p_k(\x,t|\x_0)\right ]d\x\nonumber \\
 &=-\sum_{k=1}^N \int_{\partial \calU_k}\nabla q\cdot \n_k d\sigma+\sum_{k=1}^N \int_{\partial \calU_k}\nabla p_k\cdot \n_k d\sigma  -\kappa \sum_{k=1}^N \int_{ \calU_k}p_k(\x,t|\x_0)d\x\nonumber \\
 &=-\kappa \sum_{k=1}^N \int_{ \calU_k}p_k(\x,t|\x_0)d\x,
\label{Q2}
\end{align}
where we have used the current conservation condition in equation (\ref{masterc}).
Laplace transforming equation (\ref{Q2}) and imposing the initial condition $Q(\x_0,0)=1$ gives
\begin{equation}
\label{QL}
s\widetilde{Q}(\x_0,s)-1=- \sum_{k= 1}^N \widetilde{J}_k(\x_0,s).
\end{equation}
It is also convenient to introduce the probability that the particle is captured by the $k$-th target after time $t$:
\begin{equation}\Pi_k(\x_0,t)=\int_t^{\infty}J_k(\x_0,t')dt'.
\end{equation}
In Laplace space
\begin{equation}
\label{Pi}
s\widetilde{\Pi}_k(\x_0,s)-\pi_k=-\widetilde{J}_k(\x_0,s).
\end{equation}

Now suppose that prior to being absorbed by one of the targets, the particle can instantaneously reset to a fixed location $\x_r$ at a random sequence of times generated by an exponential probability density $\psi(\tau)=r\e^{-r\tau}$, where $r$ is the resetting rate. The probability that no resetting has occurred up to time $\tau$ is then $\Psi(\tau)=1-\int_0^{\tau}\psi(s)ds=\e^{-r\tau}$. In the following we identify $\x_r$ with the initial position by setting $\x_0=\x_r$. (For simplicity we do not include resetting delays such as finite return times and refractory periods, nor non-exponential resetting statistics \cite{Chechkin18,Mendez19,Mendez19a,Evans19a,Bodrova20,Pal20,Bressloff20A}.)
Eqs. (\ref{master}a,b) become
\begin{subequations}
\label{mastera(res)}
\begin{align}
\frac{\partial q_r(\x,t|\x_0)}{\partial t} &= D\nabla^2 q(\x,t|\x_0)-rq_r(\x,t|\x_0)+r\delta(\x-\x_0),\ \x \in \calU\backslash \calU_a  ,\\
	\frac{\partial p_{r,k}(\x,t|\x_0)}{\partial t} &= D\nabla^2 p_{r,k}(\x,t|\x_0) -(\kappa+r) p_{r,k}(\x,t|\x_0),\ \x\in \calU_k.
	\end{align}
	\end{subequations}
	Here $q_r$ and $p_{r,k}$ are the resetting analogs of $q$ and $p_k$, respectively. Similarly, let $Q_r(\x_0,t)$ denote the survival probability in the presence of resetting:
\begin{align}
\label{surv}
Q_r(\x_0,t)&=\int_{\calU}p_r(\x,t|\x_0)d\x=\int_{\calU\backslash \calU_a}q_r(\x,t|\x_0)d\x+\sum_{k=1}^N \int_{\calU_k}p_{r,k}(\x,t|\x_0)d\x.\nonumber\\
\end{align} 
$Q_r$ can be related to the survival probability without resetting, $Q$, using a last renewal equation \cite{Evans11a,Evans11b,Evans20}; this holds irrespective of whether the target is partially or totally absorbing:
\begin{align}
Q_r(\x_0,t)&=\e^{-rt}Q(\x_0,t)+r\int_0^tQ(\x_0,\tau)Q_r(\x_0,t-\tau)\e^{-r\tau}d\tau.
\label{renQ}
\end{align}
The first term on the right-hand side represents trajectories with no resettings. The integrand in the second term is the contribution from trajectories that last reset at time $\tau\in (0,t)$, and consists of the product of the survival probability starting from $\x_0$ with resetting up to time $t-\tau$ and the survival probability starting from $\x_0$ without any resetting for the time interval $\tau$. 
Laplace transforming the last renewal equation and rearranging shows that
\begin{equation}
\label{Qr}
{ \widetilde{Q}_r(\x_0,s)=\frac{ \widetilde{Q}(\x_0,r+s)}{1-r \widetilde{Q}(\x_0,r+s)}.}
 \end{equation}

In the case of multiple targets with stochastic resetting, one can express the splitting probabilities and conditional MFPTs in terms of the fluxes without resetting by using Eqs (\ref{QL}), (\ref{Pi}) and (\ref{Qr}) \cite{Bressloff20A}. (Multiple targets are also considered in Ref. \cite{Chechkin18}.) Let the discrete random variable $K(t)\in \{0,1,\ldots,N\}$ indicate whether the particle has been absorbed by the $k$-th target ($K(t)=k \neq 0$) or has not been absorbed by any target ($K(t)=0$) in the time interval $[0,t]$.
The FPT that the particle is absorbed by the $k$-th target is
\begin{equation}
\calT_{r,k}=\inf\{t>0; \X(t)\in \calU_k,\ K(t)=k\},
\end{equation}
with $\calT_{r,k}=\infty$ if the particle is absorbed by another target.
The following results then hold \cite{Bressloff20A}. First,
the splitting probability that the search process with resetting finds the $k$-th target is 
\begin{align}
\pi_{r,k}(\x_0)&=\frac{\pi_{k}(\x_0)-r\widetilde{\Pi}_{k}(\x_0,r)}{ 
1-r\widetilde{Q}(\x_0,r)} =\frac{\widetilde{J}_k(\x_0,r)}{ 
\sum_{j=1}^N\widetilde{J}_j(\x_0,r)}.
\label{Piee}
\end{align}
Similarly, the Laplace transformed conditional FPT density for the $k$th target is
\begin{align}
\label{Tcond1}
 \pi_{r,k}(\x_0)\widetilde{f}_{r,k}(\x_0,s)&=\frac{\pi_{k}(\x_0)-(r+s)\widetilde{\Pi}_{k}(\x_0,r+s)}{1-r\widetilde{Q}(\x_0,r+s)}.
\end{align}
The Laplace transform of the FPT density is the moment generator of the conditional FPT:
\begin{equation}
\pi_{r,k}T_{r,k}^{(n)}=\E[{\mathcal T}_k^n1_{\Omega_k}]=\left .\left (-\frac{d}{ds}\right )^n\E[\e^{-s{\mathcal T}_k}1_{\Omega_k}]\right |_{s=0}.
\end{equation}
In particular, using equation (\ref{Piee}) and the fact that
   $ \widetilde{Q}(\x_0, r) = \sum_{k=1}^N\widetilde{\Pi}_k(\x_0, s)$, the conditional MFPT $T_{r,k}\equiv T_{r,k}^{(1)}$ is
\begin{align}
\label{Tcond2}
 & \pi_{r,k}(\x_0)T_{r,k}(\x_0)  =\frac{\widetilde{\Pi}_{k}(\x_0,r) +r\widetilde{\Pi}_{k}'(\x_0,r)-r\pi_{r,k}(\x_0)\sum_k\widetilde{\Pi}_k'(\x_0,r)}
{1-
r\widetilde{Q}(\x_0,r)} ,\end{align}
where $'$ denotes differentiation with respect to $r$.
Finally, summing over $k$ yields the unconditional MFPT
\begin{align}
\label{Ttot}
 T_{r}(\x_0)  &:=\sum_{k=1}^N \pi_{r,k}(\x_0)T_{r,k}(\x_0) =\frac{\widetilde{Q}(\x_0,r)}{1-
r\widetilde{Q}(\x_0,r)} =\frac{1-  \sum_{k= 1}^N \widetilde{J}_k(\x_0,r)}
{  r\sum_{j=1}^N\widetilde{J}_j(\x_0,r)}.
\end{align}

Equations (\ref{Piee}) and (\ref{Ttot}) imply that the splitting probabilities and unconditional MFPT with resetting are determined by the Laplace transforms of the fluxes into the targets without resetting, $\widetilde{J}_k(\x_0,r)$, $k=1,\ldots,N$. The latter depend on the absorption rate $\kappa$. In order to determine the fluxes we have to solve the Laplace transformed version of equations (\ref{master}), which take the form
 \begin{subequations}
 \label{masterLT}
\begin{align}
	&D\nabla^2 \q(\x,s|\x_0)-s\q(\x,s|\x_0)=-\delta(\x-\x_0), \ \x\in \calU \backslash \calU_a,\\
	&D\nabla^2 \p_k(\x,s|\x_0) -(\kappa+s) \p_k(\x,s|\x_0)=0,\ \x\in \calU_k,
	\end{align}
	together with the boundary condition $\nabla \q \cdot \n=0, \ \x\in \partial \calU$ and the continuity conditions 
\begin{equation}
\label{masterLTc}
 \q(\x,s|\x_0)=\p_k(\x,s|\x_0),\ \nabla \q(\x,s|\x_0)\cdot \n_k =  \nabla \p_k(\x,s|\x_0)\cdot \n_k
\end{equation}
 for all $\x \in \partial \calU_k$.
 \end{subequations}
Integrating equation (\ref{masterLT}b) over the domain $\calU_k$ implies that
\begin{align}
\label{lemmy}
(s+\kappa) \int_{\calU_k}\p_k(\x,s|\x_0)&=D\int_{\partial \calU_k}\nabla \p_k\cdot \n_k d\sigma
=
D\int_{\partial \calU_i}\nabla \q\cdot \n_k d\sigma. 
\end{align}
In the fast absorption limit $\kappa \rightarrow \infty$, the system reduces to the scalar equation
\begin{align}
\label{scal}
	&D\nabla^2 \q(\x,s|\x_0)-s\q(\x,s|\x_0)=-\delta(\x-\x_0), \ \x\in \calU \backslash \calU_a,\  \q(\x,s|\x_0)=0,\ \x \in \partial \calU_a.
\end{align}
Equation (\ref{scal}) takes the form of a classical narrow capture problem, which can be analyzed using matched asymptotic expansions and Green's function methods
\cite{Ward93,Bressloff08,Coombs09,Cheviakov11,Chevalier11,Coombs15,Ward15,Lindsay15,Lindsay17,Grebenkov20}. The basic procedure involves constructing an inner or local solution

In the following two sections we further extend the analysis to the case of partially absorbing targets as described by equations (\ref{masterLT}). We consider 2D diffusion in \S 3 and 3D diffusion in \S 4. The details of the asymptotic analyses differ due to the well known differences in the singular nature of the associated Green's function in 2D ($\ln|\x-\x'|)$ and 3D ($1/|\x-\x'|$). In particular, carrying out the matched asymptotic expansion in 2D naturally leads to terms involving the small parameter $\nu=-1/\ln \epsilon$, which is a common feature of strongly localized perturbations in 2D domains \cite{Ward93}. Since $\nu \rightarrow 0$ much more slowly than $\epsilon\rightarrow 0$, it is necessary to sum over the logarithmic terms non-perturbatively in order to obtain $O(1)$ accuracy with respect to an expansion in $\epsilon$.

\setcounter{equation}{0}
\section{Matched asymptotic analysis in 2D}

We first consider 2D diffusion and develop the asymptotic analysis by summing over all logarithmic terms, which is accurate to leading order in $\epsilon$. The corresponding analysis for totally absorbing targets was developed in \cite{Lindsay16,Bressloff21A}. The inner solution near the $j$-th target is constructed by introducing the stretched local variable ${\mathbf y} =
\varepsilon^{-1}(\x-\x_j)$ and setting
$U(\y,s|\x_0)=\q(\x_j+\varepsilon \y,s|\x_0)$ and $V(\y,s|\x_0)=\p_j(\x_j+\varepsilon \y,s|\x_0)$. In order to maintain effective absorption in the limit $\epsilon \rightarrow 0$, we also introduce the scaling $\kappa=\kappa_1/\epsilon^2$. Then $U,V$ satisfy to $O(\epsilon)$
\begin{subequations}
\label{inner0}
\begin{align}
& \nabla^2_{\y}U = 0, \   |\y| > \ell_j \quad \nabla^2_{\y}V =\frac{\kappa_1}{D} V_0, \   |\y| < \ell_j ,\\
&U(\y,s|\x_0)=V(\y,s|\x_0),\   \nabla_{\y} U\cdot \n_j = \nabla_{\y} V\cdot \n_j	 ,\ |\y |=\ell_j .
\end{align}
\end{subequations}
The solution in the original coordinates takes the form
\begin{subequations}
\begin{align}
    &U({\bf x}, s|{\bf x}_0) = A_j(\nu, s) + \nu A_j(\nu, s)\left[\Phi(\beta\ell_j) + \log|{\bf x} - {\bf x}_j|/\ell_j\right],\\
    &V(\x, s|\x_0) = \frac{\nu A_j(\nu, s)\Phi(\beta\ell_j)}{I_0\left(\beta\ell_j\right)}I_0\left(\beta\epsilon^{-1}|\x - \x_j|\right)
\end{align}
\end{subequations}
where $\nu=-1/\ln \epsilon$,
\begin{equation}
\label{Phi}
\Phi  = \frac{I_0(\beta\ell_j)}{\beta\ell_j  I_1(\beta\ell_j)},\quad  \beta = \sqrt{\kappa_1/D},
\end{equation}
and $I_n(y)$ is a modified Bessel function of the first kind.

The outer solution is constructed by shrinking each target to a single point and imposing a corresponding singularity condition that is obtained by matching with the inner solution. The outer equation is given by
\begin{align}
\label{outer}
	 D\nabla^2 \q(\x,s|\x_0) -s\q(\x,s|\x_0)= -\delta(\x-\x_0) , \ 
		\end{align}
for $\x\in \calU\backslash \{\x_1,\ldots,\x_N\},$ together with the boundary condition $\nabla \q \cdot \n=0, \ \x\in \partial \calU$. The corresponding singularity conditions are
\begin{align}
    \q \sim A_j(\nu, s) + \nu A_j(\nu, s)\left[\Phi(\beta\ell_j) + \log|{\bf x} - {\bf x}_j|/\ell_j\right]
\end{align}
for ${\bf x} \to {\bf x}_j$. The next step is to introduce the Green's function of the 2D modified Helmholtz equation according to
\begin{subequations}
\begin{align}
\label{GMH}
	&-\delta(\x-\x_0) = D\nabla^2 G(\x,s|\x_0) -sG(\x,s|\x_0), \ \x\in \calU,\\
	&0 = \nabla G(\x,s|x_0)\cdot \n ,\ \x \in \partial \calU,\int_{\calU} G(\x,s|x_0)d\x=-\frac{1}{s} . 
	\end{align}
	\end{subequations}
Note that $G$ can be decomposed as 
\begin{align}
    G(\x, s; \x_0) = -\frac{\log{|\x - \x_0|}}{2\pi D} + R(\x, s; \x_0),
\end{align}
where $R$ is the non-singular part of the Green's function. We then set
\begin{equation}
\q(\x,s|\x_0) = G(\x,s|x_0) + \psi(\x, s),\ 
\end{equation}
such that
\begin{align}
\label{psi}
    D\nabla^2\psi(\x, s) - s\psi(\x, s) = 2\pi\nu D\sum_{j=1}^nA_j(\nu, s)\delta(\x - \x_j)
\end{align}
for $\x \in \calU .$ This is supplemented by the boundary condition $\nabla \psi\cdot \n = 0$ on $\partial \calU$. It follows that $\psi$ has a solution of the form
\begin{align}
    \psi(\x, s) = - 2\pi\nu D\sum_{j=1}^n A_j(\nu, s)G(\x,s|x_j).
\end{align}

We have $N$ unknown coefficients $A_j(\nu,s)$, which are obtained by solving $N$ constraints obtained by matching the inner and outer solutions:
\begin{align}
     G({\bf x}_j, s|{\bf x}_0) &= A_j(\nu, s)\left[1 + \nu\Phi(\beta\ell_j) + 2\pi\nu DR({\bf x}_j,s |{\bf x}_j)\right]\nonumber \\\
     &\quad + 2\pi\nu D\sum_{i \neq j}A_i(\nu, s)G({\bf x}_i, s| {\bf x}_j) .
  \label{match}
\end{align}
These $N$ equations can be represented by the matrix equation
\begin{align}
\label{matrix}
    \left[\left(1 + \nu\Phi(\beta\ell_j)\right){\bf I} + 2\pi \nu D{\bf {\bf G}^{\top}}\right]{\bf a} = {\bf g},
\end{align}
where ${\bf I}$ is the $N \times N$ identity matrix,
\begin{align}
    {\bf a}&=(A_1(\nu, s), \ldots, A_N(\nu, s))^T, \ {\bf g} = (G({\bf x}_1, s| {\bf x}_0), \ldots, G({\bf x}_N, s|{\bf x}_0))^T,
\end{align}
and ${\bf G}$ is an $N \times N$ matrix given by
\begin{align}
    {\bf G}_{jj} = R({\bf x}_j, s| {\bf x}_j), \quad {\bf G}_{ij} = G({\bf x}_i, s| {\bf x}_j), \ j \neq i.
\end{align}
Inverting equation (\ref{matrix}) yields in component form
for $j = 1, \ldots, N$,
\begin{align}
\label{naj}
    A_j(\nu, s) 
    &= \sum_{i=1}^N\left[\left(1 + \nu\Phi(\beta\ell_j)\right){\bf I} + 2\pi \nu D{\bf {\bf G}^{\top}}\right]^{-1}_{ji}G(\x_i, s|\x_0).
    \end{align}
    This is a non-perturbative solution that sums over all logarithmic terms along analogous lines to \cite{Ward93}.

 The coefficients $A_j$ of equation (\ref{Aj}) determine the splitting probabilities and conditional MFPTs under resetting. Using equation (\ref{lemmy}) we can write
\begin{align}
   \widetilde{J}_i(\x_0, s) =D\frac{\kappa_1}{\epsilon^2s+\kappa_1}\int_{\partial \calU_i}\nabla \q\cdot \n d\sigma ,
    \end{align}
where $\q$ is the inner solution (\ref{inner0}a) so that
\begin{align*}
 \left.\nabla_{\x}\q \cdot {\bf n}\right|_{\x \in \partial\calU_j} = \epsilon^{-1}\left.\nabla_{{\bf y}}U_0 \cdot {\bf n}\right|_{|{\bf y}|=\ell_j} \sim \epsilon^{-1}\ell_j^{-1}\nu A_j(\nu, s),   
\end{align*}
Hence, to $O(1)$ in $\epsilon$
\begin{align}
   \widetilde{J}_i(\x_0, s) \sim  2\pi \nu DA_i(\nu, s) .
\end{align}
Substituting for the flux  into Eqs. (\ref{Piee}) and (\ref{Ttot}) for the hitting probability $\pi_{r,k}(\x_0)$ and unconditional MFPT $T_r(\x_0)$ with resetting gives
\begin{align}
\label{TtotN}
 \pi_{r,k}(\x_0)  &=\frac{A_k(\nu, r)}
{\sum_{j=1}^NA_j(\nu, r)},\quad
 T_{r}(\x_0)  =\frac{1-2\pi \nu D \sum_{k= 1}^N A_k(\nu, r)}
{2\pi \nu Dr\sum_{j=1}^N A_j(\nu, r)}.
\end{align}

 It follows from the above analysis that all information regarding absorption within a target domain $\calU_j$, including the effective absorption rate $\kappa_1$, is contained in the function $\Phi(\beta\ell_j)$ defined in (\ref{Phi}) with $\beta=\sqrt{\kappa_1/D}$. In particular, $\Phi(\beta\ell_j)$ is an exponentially decreasing function of $\beta$ with $\Phi(\beta\ell_j)\rightarrow 0$ as $\beta\rightarrow \infty$ and $\Phi(\beta\ell_j)\rightarrow \infty$ as $\beta\rightarrow 0$. It immediately follows that the results for totally absorbing targets \cite{Bressloff21A} are recovered  in the fast absorption limit. On the other hand, $A_j(\nu,s)\rightarrow 0$ in the limit $\kappa_1\rightarrow 0$. The latter means that the net flux into each target is zero since there is no absorption.
Hence, the dependence of $T_r$ on $\Phi(\beta\ell_j)$ implies that $T_r(\x_0)\rightarrow \infty$ as $\kappa_1\rightarrow 0$, whereas it converges to the unconditioned MFPT for totally absorbing target boundaries when $\kappa_1\rightarrow \infty$.

\subsection{Example}

In order to determine the coefficients $A_k(\nu, s)$ we need to obtain accurate numerical or analytical approximations of the Green's function for the modified Helmholtz equation and solve the matrix equation (\ref{matrix}). This particular issue has been addressed by Lindsay {\em et al.}  \cite{Lindsay16},whose results can be applied to the current problem. An important step in the evaluation of the Green's function is to decompose $G$ as the sum of the free-space Green's function and a regular boundary-dependent part:
\begin{align}
    G(\x, s;|\x_0) = \frac{1}{2\pi D}K_0\left(\sqrt{s / D}|\x - \x_0|\right) + \widehat{R}(\x, s|\x_0)
\end{align}
where $K_0$ is the modified Bessel function of the second kind and $\widehat{R}$ is non-singular at $\x = \x_0$.  It can be shown that for $|\x - \x_0| = O(1)$ and large $\sqrt{s / D}$,  the boundary contributions to $\widehat{R}$ are exponentially small.  This allows us to write
\begin{align*}
    G(\x, s;|\x_0) &\sim \frac{1}{2\pi D}K_0\left(\sqrt{s / D}|\x - \x_0|\right), \ \x \neq \x_0, \\
  \widehat{R}(\x_0, s;|\x_0) &\sim -\frac{1}{2\pi D}\left(\ln\sqrt{s / D} - \ln2 + \gamma_c\right)
\end{align*}
where $\gamma_c \approx 0.5772$ is Euler's gamma constant. It follows that the off-diagonal terms in equation (\ref{matrix}) are exponentially smaller than the diagonal terms. Therefore, we have to leading order
\begin{align}
\label{Aj}
   A_j(\nu, s) \sim \frac{(2\pi D)^{-1}K_0\left(\sqrt{s/D}|\x_j - \x_0|\right)}{1 + \nu\Phi(\beta\ell_j) - \nu\left[\ln\sqrt{s / D} - \ln2 + \gamma_c\right]}. 
\end{align}
Hence, under the boundary-free approximation, $A_j(\nu,s)$ depends on the distances of the targets from $\x_0$ but is independent of the shape of the domain and the absolute locations of the targets. Numerically it has been shown that such an approximation remains valid even at intermediate values of $s$ (or equivalently at intermediate times) provided that $\x_0$ and $\x_j$, $j=1,\dots,N$, are not close to the boundary and there are no bottlenecks separating the targets from $\x_0$. This result is reinforced in the presence of resetting.
Hence, under the further approximation (\ref{Aj}) and assuming $\ell_j=\ell$ for all $j$, equations (\ref{TtotN}) become
\begin{align}
    &\pi_{r, j}(\x_0) \sim \frac{ K_0\left(\sqrt{r/D}|\x_j - \x_0|\right)}{\sum_{k=1}^NK_0\left(\sqrt{r/D}|\x_k - \x_0|\right)},\\
    \label{mfpt-2d}
    T_r(\x_0) &\sim \frac{1}{r}\frac{1 + \nu\Phi(\beta\ell) - \nu\left[\ln\sqrt{r / D} - \ln2 + \gamma_c\right] - \nu\sum_{k=1}^NK_0\left(\sqrt{r/D}|\x_k - \x_0|\right)}{\nu\sum_{k=1}^NK_0\left(\sqrt{r/D}|\x_k - \x_0|\right)}.\nonumber \\
\end{align}

\begin{figure}[t!]
    \centering
    \includegraphics[width=13cm]{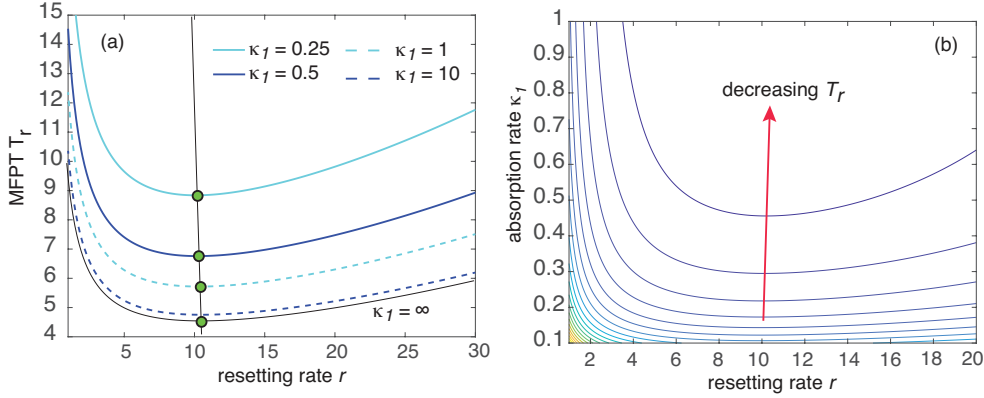}
    \caption{Single small target in $\R^2$ with $|\x_1-\x_0|\equiv \rho_1=0.5$. (a) Plot of the MFPT $T_r(\x_0)$ vs the resetting rate $r$ for various absorption rates $\kappa_1$. Other parameters are $\nu=0.1$, $\ell=1$ and $D=1$. Filled circles indicate the optimal resetting rate. (b) Corresponding contour plot.}
    \label{fig3}
\end{figure}

\begin{figure}[b!]
    \centering
    \includegraphics[width=13cm]{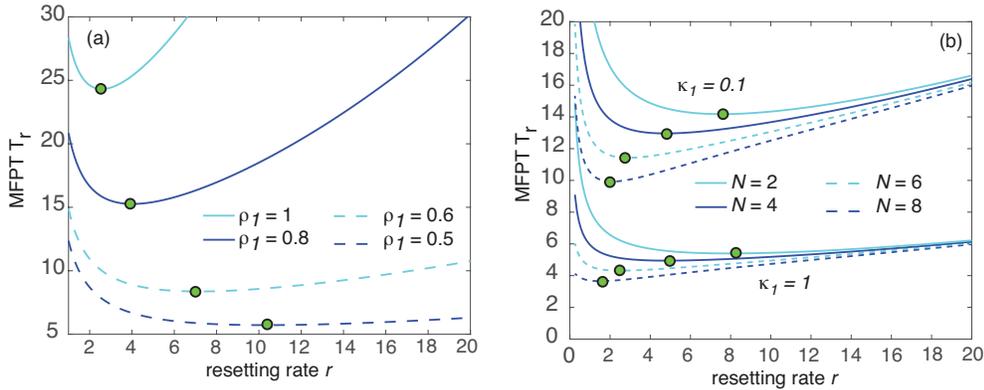}
    \caption{(a) Same as Fig. \ref{fig3} for different target distances $\rho_1$ and $\kappa_1=1$. (b) $N+1$ targets in $\R^2$ with $|\x_j - \x_0| = \rho_0 + (j - 1)\Delta\rho$ for $j=1,\ldots,N$. Unconditional MFPT $T_r(\x_0)$ vs resetting rate for fixed $\kappa_1=0.1,1$. Other parameter values are $\nu=0.1$, $\ell=1$, $\rho_0 = 0.5$, $\Delta\rho = 1$, and $D=1$. Filled circles indicate the optimal resetting rate.}
    \label{fig4}
\end{figure}

In Fig. \ref{fig3}(a) we plot $T_r$ as a function of $r$ for a single target with $|\x_1-\x_0| \equiv \rho_1=0.5$. A corresponding contour plot is shown in Fig. \ref{fig3}(b). It can be seen that $T_r$ has a minimum at an optimal resetting rate $r_{\rm opt}$, which is only weakly dependent on $\kappa_1$. As expected, increasing $\kappa_1$ reduces $T_r$ as the particle has a higher probability for absorption. We also note that the MFPT is most sensitive to variations in $\kappa_1$ at small resetting rates.
 In all cases, $T_r\rightarrow \infty$ as $r\rightarrow \infty$, which reflects the fact that if the particle resets too often then it never has the chance to reach even the closest target. The divergence of $T_r$ can be explored further using the asymptotic expansion
\begin{align}
K_0(z)&\sim \sqrt{\frac{\pi}{2  z}}\e^{-z}\left [1-\frac{1}{8  z}+O( z^{-2})\right ],\quad z\rightarrow \infty,
\end{align}
which implies that
\begin{equation}
\label{eqT}
 T_r(\x)\sim (1+\nu \Phi(\beta\ell ))\frac{\sqrt{\sqrt{r /D}|\x_1-\x_0}|}{\nu r}\sqrt{\frac{2}{\pi}} \e^{\sqrt{r/D}|\x_1-\x_0|}
\end{equation}
as $ r\rightarrow \infty$. In contrast to its dependence on $\kappa_1$, the optimal resetting rate increases significantly with the target distance $\rho_1$, as shown in Fig. \ref{fig4}(a).
Analogous results hold for multiple target configurations as illustrated in Fig. \ref{fig4}(b) for a set of $N+1$ targets whose distances from $\x_0$ are of the form $\rho_j=|\x_j-\x_0| =\rho_0+(j-1)\Delta \rho$, $j=1,\ldots,N$.

\section{Matched asymptotics in 3D}

We now turn to the corresponding asymptotic analysis of the 3D problem, following along the lines of  \cite{Cheviakov11,Coombs15,Bressloff21B} for totally absorbing targets. The outer solution for $\q(\x,s|x_0)$ is expanded as
\[\q(\x,s|\x_0)\sim \q_0(\x,s|\x_0)+\epsilon\q_1(\x,s|\x_0)+\epsilon^2 \q_2(\x,s|\x_0)+\ldots
\]
The leading order term $\q_0$ satisfies the Laplace transformed diffusion equation without any targets within $\calU$:
\begin{align}
\label{asym0}
D\nabla^2 \q_0-s\q_0&=-\delta(\x-\x_0),\, \x\in \calU;\ \nabla \q_0\cdot \n=0,\, \x\in \partial \calU.
\end{align}
That is, $\q_0=G(\x,s|\x_0)$
where $G$ is now the Neumann Green's function of the 3D modified Helmholtz equation. The latter can be decomposed as
\begin{align}
\label{GR}
	 G(\x,s|\x_0)&=\frac{1}{4\pi D|\x-\x_0|}+R(\x,s|\x_0),
	\end{align}
where $R$ is the regular (non-singular) part of $G$. The higher-order contributions to the outer solution satisfy the equations
\begin{align}
\label{asym1}
D\nabla^2 \q_n-s\q_n&=0,\, \x\in \calU\backslash \{\x_1,\ldots,\x_N\};\ \nabla \q_n\cdot \n=0,\, \x\in \partial \calU,
\end{align}
for $n \geq 1$, together with certain singularity conditions as $\x\rightarrow \x_j$, $j=1,\ldots,N$, which are determined by matching to the inner solution.

The inner solution near $\calU_j$ is constructed in terms of the stretched local variable ${\bf y}=\epsilon^{-1}(\x-\x_j)$ by setting
$U(\y,s|\x_0)=\q(\x_j+\varepsilon \y,s|\x_0)$ and $V(\y,s|\x_0)=\p_j(\x_j+\varepsilon \y,s|\x_0)$, see Fig. \ref{fig3}. Then $U,V$ satisfy
\begin{subequations}
\begin{align}
\label{innerA}
& D\nabla^2_{\y}U   -s\epsilon^2 U=0,\  \   |\y| > \ell_j \quad D\nabla^2_{\y}V  -(s+\kappa)\epsilon^2 V =0, \   |\y| < \ell_j ,\\
&U(\y,s|\x_0)=V(\y,s|\x_0),\quad   \nabla_{\y} U\cdot \n_j = \nabla_{\y} V\cdot \n_j	 ,\quad |\y |=\ell_j .
\label{innerB}
\end{align}
\end{subequations}
Now consider a perturbation expansion of the inner solution around the $j$-th target of the form 
\[ U\sim U_0 + \epsilon U_1 + \epsilon^2U_2+O(\epsilon^3),\quad V\sim V_0 + \epsilon V_1 + \epsilon^2V_2+O(\epsilon^3).
 \]
 Assuming $\kappa=\kappa_1/\epsilon^2$ and $s\ll 1/\epsilon$, this yields the hierarchy of equations
\begin{subequations} 
\label{stretch}
\begin{align}
&D\nabla_{\y}^2 U_n(\y,s) =0,\  |\y|>\ell_j,\quad D\nabla_{\y}^2 V_n(\y,s) -\kappa_1 V_n=0,\ |\y|<\ell_j,\ n=0,1\\
	&D\nabla_{\y}^2 U_n(\y,s) =s U_{n-2}(\y,s), \  \  |\y|>\ell_j \ n\geq 2,\\
&	D\nabla^2_{\y}V_n  -\kappa_1V_n= sV_{n-2}, \ |\y|<\ell_j, \ n\geq 2,\\
&	 U_n(\y,s)= V_n(\y,s),\quad  \nabla_{\y} U_n\cdot \n_j = \nabla_{\y} V_n\cdot \n_j, \  |\y|=\ell_j,\ n\geq 0.
	\end{align}
\end{subequations}
These are supplemented by far-field conditions obtained by matching $U$ with the near-field behavior of the outer solution:
\begin{equation}
U_0 + \epsilon U_1 + \epsilon^2U_2+\ldots \sim   \q_0 + \epsilon \q_1 + \epsilon^2\q_2+\ldots
 \end{equation}

The matching is developed iteratively along the lines of \cite{Bressloff21B}, starting from the known leading order contribution to the outer solution. Taylor expanding the latter near
the $j$-th region $\calU_j$ yields 
\begin{equation}
\q_0 \sim G(\x_j,s|\x_0) + \nabla_{\x} G(\x,s|\x_0)\vert_{\x=\x_j} \cdot (\x-\x_j)+\frac{1}{2}{\bf H}_j\cdot(\x-\x_j)\otimes (\x-\x_j)+\ldots,
\end{equation}
where ${\bf H}_j $ is the Hessian
\begin{equation}
H_j^{ab}=\left . \frac{\partial^2}{\partial x_a\partial x_b}G(\x,s|\x_0)\right \vert_{\x=\x_j} ,\quad a,b\in \{1,2,3\}.
\end{equation}
In terms of the stretched coordinate ${\bf y}$, we have
\begin{equation}
\label{p0}
\q_0 \sim G(\x_j,s|\x_0)+ \epsilon \nabla_{\x} G(\x_j,s|\x_0) \cdot \y +\frac{ \epsilon^2}{2} {\bf H}_j \cdot \y\otimes \y+\ldots
\end{equation}
It follows that the $n$-th order derivative of $G$ will contribute to the far-field behavior of $U_n$, along with lower order derivatives of the the $1,\dots, n-1$ terms. Additional contributions will arise from the non-singular terms in $\q_n$. On the other hand, the inner solution $U_n$ will have a term proportional to $1/|\y|=\epsilon/|\x-\x_j|$ that determines the singular behavior of $\q_{n+1}$ as $\x\rightarrow \x_j$. We will use this iterative matching procedure to calculate the terms $U_0,U_1,U_2$, and then use these to obtain the target fluxes to $O(\epsilon^3)$. The major difference from our previous study of the 3D narrow capture problem \cite{Bressloff21B} is that we also have to match $U_n$ with the solution $V_n$ inside a target; this is crucial for quantifying how the fluxes depend on the absorption rate $\kappa_1$.

\subsection{Calculation of the leading order inner solution $U_0$}

Setting $n=0$ in equations (\ref{stretch}) and matching the far-field behavior of $U_0$ with the near-field behavior of $\q_0$ leads to the set of equations
\begin{subequations} 
\begin{align}
D\nabla_{\y}^2 U_0(\y,s) &=0,\ |\y|>\ell_j,U_0 \sim G(\x_j,s|\x_0) \mbox{ as } |\y|\to \infty, \\
D\nabla_{\y}^2 V_0(\y,s) -\kappa_1 V_0&=0,\quad  |\y|<\ell_j,   \\
U_0(\y,s)=& V_0(\y,s),\quad  \nabla_{\y} U_0\cdot \n_j = \nabla_{\y} V_0\cdot \n_j, \  |\y|=\ell_j.
\end{align}
\end{subequations}
The general solution is of the form
\begin{equation}
\label{gen0}
U_0 = G(\x_j,s|\x_0)+{A_0}\frac{\ell_j}{|\y|},\quad V_0=B_0\frac{I_{1/2}(\beta|\y|)}{\sqrt{\beta|\y|}}, \quad \beta=\sqrt{\frac{\kappa_1}{D}}.
\end{equation}
Note that we can write $I_{1/2}(x) = \sqrt{2/\pi}\sinh(x) / \sqrt{x}$ so that $V_0 = B_0\sinh{(\beta|\y|)}/\beta |\y|$.
The matching conditions then imply that
\begin{subequations}
\begin{align}
 &G(\x_j,s|\x_0)+{A_0}=\frac{B_0}{\sqrt{\beta \ell_j}}I_{1/2}(\beta\ell_j),\\ -\frac{A_0}{\ell_j}&=\frac{B_0}{\sqrt{\beta \ell_j}}\left [\beta I'_{1/2}(\beta\ell_j)  -\frac{1}{2\ell_j} I_{1/2}(\beta\ell_j) \right ]=\frac{B_0}{\beta \ell_j^2}\left [\beta\ell_j \cosh(\beta\ell_j) - \sinh(\beta\ell_j) \right ].
\end{align}
\end{subequations}
Hence,
\begin{equation}
U_0 = G(\x_j,s|\x_0)\left [1-\Phi(\beta\ell_j) \frac{\ell_j}{|\y|}\right ],
\end{equation}
where 
\begin{equation}
\label{phib}
\Phi(\beta\ell_j)=\frac{2\ell_j \beta I'_{1/2}(\beta\ell_j)-I_{1/2}(\beta\ell_j) }{2\ell_j \beta I'_{1/2}(\beta\ell_j)+I_{1/2}(\beta\ell_j) }=1 - \frac{\tanh(\beta \ell_j)}{\beta\ell_j} .
\end{equation}
It now follows that $\q_1$ satisfies equation (\ref{asym1}) together with the singularity condition
\begin{equation}
\q_1(\x,s)\sim - \Phi(\beta\ell_j)\frac{G_{j0}\ell_j}{|\x-\x_j|} \quad \mbox{as } \x\rightarrow \x_j,
\end{equation}
For notational convenience, we have set $G_{j0}=G(\x_j,s|\x_0)$ and dropped the explicit dependence on $s,\x_0$. Hence, $\q_1(\x,s)$ satisfies the inhomogeneous equation
\begin{align}
 D\nabla^2 \q_1-s\q_1&={4\pi D}\sum_{j=1}^N \Phi(\beta\ell_j)G_{j0}\ell_j \delta(\x-\x_j),\, \x\in \calU;\quad
 \nabla \q_1\cdot \n=0,\ \x \in \partial \calU,
 \label{asym2}
\end{align}
which can be solved in terms of the modified Helmholtz Green's function:
\begin{equation}
\label{q1}
\q_1(\x,s)=- {4\pi}  D\sum_{j=1}^N \Phi(\beta\ell_j)G_{j0}\ell_jG(\x,s|\x_j).
\end{equation}

\subsection{Calculation of higher-order terms} The next step is to match the far-field behavior of $U_1$ with the $O(\epsilon)$ term in the expansion of $\q_0$, see equation (\ref{p0}) and the near field behavior of $\q_1 $ around $\calU_j$. The latter takes the form
\begin{align*}
\q_1(\x,s)&= - \Phi(\beta\ell_j)\bigg\{\frac{G_{j0}\ell_j}{|\x-\x_j|}+ {4\pi}D G_{j0} \ell_jR(\x_j,s|\x_j) \bigg \}\\
&\quad -4\pi D\sum_{k\neq j}^N \Phi(\beta\ell_k)G_{k0}\ell_kG(\x_j,s|\x_k)
\end{align*}
It follows that $U_1$ is determined by equations (\ref{stretch}) for $n=1$, supplemented by the condition
\begin{align}
U_1(\y,s)\rightarrow  \nabla_{\x} G(\x_j,s|\x_0) \cdot \y -4\pi D\sum_{k=1}^N \Phi(\beta\ell_k)G_{k0}\ell_k{\mathcal G}_{jk} \mbox{ as } |\y|\rightarrow \infty,
\end{align}
where
${\mathcal G}_{ij} =G(\x_i,s|\x_j)$ for $i\neq j$, and ${\mathcal G}_{ii} =R(\x_i,s|\x_i)$. It is convenient to introduce the decomposition $U_1=W_1+\widehat{W}_1$ with $\widehat{W}_1=0$ on $|\y|=\ell_j$,
\begin{equation}
\label{chi1}
W_1\rightarrow \chi_j^{(1)}=-4\pi D \sum_{k=1}^N \Phi(\beta\ell_k)G_{k0} \ell_k{\mathcal G}_{jk} \mbox{ as } |\y|\rightarrow \infty
\end{equation}
and
\begin{equation}
\widehat{W}_1\rightarrow {\bf b}_j\cdot \y \mbox{ as } |\y|\rightarrow \infty,\ {\bf b}_j= \nabla_{\x} G(\x_j,s|\x_0) .
\end{equation}
The general solutions for $W_1$ and $V_1$ are
\begin{equation}
W_1= \chi_j^{(1)}+{A_1}\frac{\ell_j}{|\y|},\quad V_1= B_1I_{1/2}(\beta|\y|)/\sqrt{\beta|\y|}.
\end{equation}
In terms of spherical polar coordinates $\rho=|\y|$, ${\bf b}_j=(0,0,b_j)$ and $\y\cdot {\bf b}_j=b_j\rho\cos \theta$, $0\leq \theta \leq \pi$, we have
\begin{align}
&\frac{\partial^2\widehat{W}_1}{\partial \rho^2}+\frac{2}{\rho}\frac{\partial \widehat{W}_1}{\partial \rho} +\frac{1}{\rho^2\sin \theta}\frac{\partial }{\partial \theta}\left (\sin \theta\frac{\partial \widehat{W}_1}{\partial \theta} \right )=0,\  \rho>1,\\
&   \widehat{W}_1 \sim b_j\rho \cos \theta  \mbox{ as } \rho \to \infty;\quad \widehat{W}_1=0 \mbox{ on } \rho=\ell_j. 
\end{align}
The general solution of Laplace's equation in spherical polar coordinates is given by
\begin{equation}
\label{gen}
B(\rho,\theta,\phi) = \sum_{l\geq 0}\sum_{m=-l}^l \left (a_{lm}\rho^l+\frac{b_{lm}}{\rho^{l+1}}\right )P_l^m(\cos \theta)\e^{im\phi},
\end{equation}
where $P_l^m(\cos \theta)$ is an associated Legendre polynomial. Imposing the Dirichlet boundary condition and the far-field condition implies that
\begin{equation}
\label{B1}
\widehat{W}_1(\y)=b_j\ell_j\cos \theta \left (\frac{|\y|}{\ell_j}-\frac{\ell_j^2}{|\y|^2}\right ).
\end{equation}
This will contribute to the far-field behavior of the $O(\epsilon^3)$ term in the outer solution. However, it does not contribute to the flux into a target.
Also note that
\begin{equation}
 \nabla \widehat{W}_1(\y)\cdot \n=3b_j\cos \theta \quad \mbox{on } |\y|=\ell_j.
 \end{equation}
We can now determine the unknown coefficients $A_1,B_1$ using the matching conditions
\begin{align}
W_1(\y)=V_1(\y),\quad \nabla W_1(\y)\cdot \n+ \nabla \widehat{W}_1(\y)\cdot \n_j =\nabla V_1(\y)\cdot \n_j ,\quad |\y|=\ell_j.
\end{align}
The matching conditions then imply that
\begin{subequations}
\begin{align}
 &\chi_j^{(1)}+{A_1}=\frac{B_1}{\sqrt{\beta \ell_j}}I_{1/2}(\beta\ell_j),\\ -\frac{A_1}{\ell_j} +3b_j\cos \theta&=\frac{B_1}{\sqrt{\beta \ell_j}}\left [\beta I'_{1/2}(\beta\ell_j)  -\frac{1}{2\ell_j} I_{1/2}(\beta\ell_j) \right ].
\end{align}
\end{subequations}
Hence,
\begin{equation}
\label{P1}
W_1 =\chi_j^{(1)}\left [1-\Phi(\beta\ell_j) \frac{\ell_j}{|\y|}\right ]+3b_j\cos \theta [1-\Phi(\beta\ell_j)] \frac{\ell_j^2}{|\y|}.
\end{equation}

The analysis of $U_n$ for $n\geq 2$ is more complicated due to the fact that the equation for $V_n$, equation (\ref{stretch}c), is now inhomogeneous. The case $n=2$ is given in the appendix. One finds that $U_n$ has the general form 
\begin{align}
U_n(\y)&=\chi_j^{(n)}-(\Phi(\beta\ell_j) \chi_j^{(n)}-w_j^{(n)})\frac{\ell_j}{|\y|}+\widehat{w}_j^{(n)}(|\y|)) \nonumber \\
&\quad +\mbox{higher-order spherical harmonics},
\label{ugh}
\end{align}
where
the coefficients $\chi_j^{(n)}$ satisfy the iterative equation
\begin{equation}
\label{chin}
\chi_j^{(n+1)}=-4\pi D\sum_{k=1}^N \Phi(\beta\ell_k)\chi_k^{(n)}\ell_k{\mathcal G}_{jk},\quad n\geq 1.
\end{equation}
Note that $w_j^{(0)}=\widehat{w}_j^{(0)}=0$, whereas equations (\ref{B1}) and (\ref{P1}) yield the expressions $w^{(1)}_j = b_j\ell_j\cos(\theta)(1 - \Phi(\beta\ell_j))$ and $\widehat{w}^{(1)}_j = b_j\ell_j\cos(\theta)\left(|\y|/\ell_j - \ell_j^2 / |\y|^2\right)$. The calculations of $w_j^{(n)}$ and $\widehat{w}_j^{(n)}(\y)$ are carried out in the appendix.

\subsection{The flux}

The $\epsilon$ expansion of the inner solution generates a corresponding expansion of the flux into the $j$-th target. In particular, Laplace transforming equation (\ref{lemmy}) and rescaling by the factor $\kappa_1/(\kappa_1+\epsilon^2s)$ shows that
\begin{align}
\widetilde{J}_j(\x_0,s)&= D\epsilon^2 \int_{|\y|=\ell_j} \nabla_{\x} U \cdot \n_j \ d\y\sim 
D \epsilon  \int_{|\y|=\ell_j}\left [ \nabla_{\y} U_0+\epsilon  \nabla_{\y} U_1+\ldots \right ] \cdot \n_j \ dS_{\y}.
\end{align}
Introducing spherical polar coordinates $(\rho,\theta,\phi)$ relative to the center of $\calU_j$, we can rewrite the asymptotic expansion of the flux as
\begin{align}
\label{JLTn}
\widetilde{J}_j(\x_0,s)&\sim \epsilon \widetilde{J}_j^{(0)}(\x_0,s)+\epsilon^2 \widetilde{J}_j^{(1)}(\x_0,s)+\epsilon^3 \widetilde{J}_j^{(2)}(\x_0,s)+\ldots
\end{align}
with
\begin{align}
\widetilde{J}_j^{(n)}&= D  \ell_j^2 \int_0^{2\pi}\int_0^{\pi} \left . \frac{\partial}{\partial \rho}\right \vert_{\rho= \ell_j} \left [-\frac{  \ell_j[\Phi(\beta\ell_j) \chi_j^{(n)}-w_j^{(n)}]}{\rho}+\widehat{w}_j^{(n)}(\rho)\right ]\sin\theta d\theta d\phi \nonumber\\
&=4\pi D\ell_j [\Phi(\beta\ell_j) \chi_j^{(n)}-w_j^{(n)}]+4\pi D\ell_j^2  \left . \frac{d \widehat{w}_j^{(n)}}{d\rho}\right |_{\rho=\ell_j}.
\end{align}
In particular, the flux through the $j$-th target is
\begin{align}
\widetilde{J}_j(\x_0,s)
&\sim 4\pi \epsilon D\ell_j\Phi(\beta \ell_j)\left (G(\x_j,s|\x_0)-4\pi \epsilon D\sum_{k=1}^N\Phi(\beta \ell_k)G(\x_k,s|\x_0)\ell_k{\mathcal G}_{jk}(s)\right )\nonumber \\
&\quad +O(\epsilon^3).
\label{JLTr}
\end{align}
This expansion will be valid provided that $s\ll 1/\epsilon$. Several remarks are in order.
\medskip

\noindent (i) The series expansion (\ref{JLTr}) to $O(\epsilon^2)$ has a relatively straightforward dependence on the rescaled absorption rate $\kappa_1$. That is, each Green's function product of order $n$ is scaled by a factor of $\Phi(\beta\ell_j)^n$, where $\Phi(\beta \ell_j)$ is defined in equation (\ref{phib}) with $\beta=\sqrt{\kappa_1/D}$. However, there are additional terms at higher-orders in $\epsilon$ that display a more complicated dependence on $\kappa_1$ due to the coefficients $w_j^{(n)}$ and $\widehat{w}_j^{(n)}$. 
\medskip

\noindent (ii) Since $\Phi(\beta\ell_j)\rightarrow 1$ in the limit $\kappa_1\rightarrow \infty$, we recover our previous result for totally absorbing targets \cite{Bressloff21B}. On the other hand, $\Phi(\beta\ell_j)\rightarrow 0$ as $\kappa_1\rightarrow 0$, which means that the net flux into each target vanishes in the absence of any absorption.
\medskip

\noindent (iii) Our previous analysis of the 3D narrow capture problem with totally absorbing targets established that the limit $s \rightarrow 0$ in equation (\ref{JLTr}) is non-trivial due to the small-$s$ singularity of the modified Helmholtz Green's function \cite{Bressloff21B}:
\begin{equation}
\label{GGG}
G(\x,s|\x')=\frac{1}{s|\calU|}+\overline{G}(\x,\x')+s {F}(\x,\x')+O(s^2),
\end{equation}
where $\overline{G}(\x,\x')$ is the Neumann Green's function for the diffusion equation:
\begin{subequations}
\label{G0}
\begin{align}
D\nabla^2 \overline{G}(\x;\x')=\frac{1}{|\calU|} -\delta(\x-\x'),\, \x\in \calU;\ \nabla \overline{G}\cdot \n=0,\, \x \in \partial \calU,\\
\overline{G}(\x,\x')=\frac{1}{4\pi D|\x-\x'|}+\overline{R}(\x,\x'),\quad \int_{\calU}\overline{G}(\x,\x')d\x=0,
\end{align}
\end{subequations}
and $\overline{R}(\x,\x')$ is regular part of $\overline{G}(\x,\x')$. 
Substitution of equation (\ref{GGG}) into equation (\ref{JLTr}) leads to terms involving factors of $\epsilon/r$, which become arbitrarily large as $r\rightarrow 0$, thus leading to a breakdown of the $\epsilon$ expansion. Following along analogous lines to \cite{Bressloff21B}, it is possible to perform a partial resummation of the asymptotic expansion that renders the resulting series non-singular in the limit $r\rightarrow 0$. The basic idea is to introduce a new dimensionless parameter
\begin{equation}
\Lambda = \frac{ 4\pi \epsilon  D \bar{\ell}}{r|\calU|}.
\end{equation}
This converts a subset of $O(\epsilon^n)$ terms in the expansion of $\widetilde{J}_j$ to $O(\epsilon^{r} \Lambda^{n-r})$ terms, $0\leq r\leq n$. At each order of $\epsilon$, we obtain an infinite power series in $\Lambda$ that can be summed to remove all singularities in the limit $r\rightarrow 0$. Following similar steps to \cite{Bressloff21B}, we find that
\begin{align}
\label{ee}
\widetilde{J}_j(\x_0,r)
&\sim 4\pi \epsilon D\ell_j \Phi(\beta\ell_j)\overline{G}_{j0}  + \frac{\ell_j}{\bar{\ell}}\frac{\Lambda}{1+\Lambda}\bigg [ 1-4\pi \epsilon D\Phi(\beta\ell_j) \sum_k\Phi(\beta\ell_k)\ell_k (\overline{G}_{k0}+\overline{\mathcal G}_{jk})\bigg ]\nonumber\\
&\quad +\frac{4\pi \epsilon \ell_j D\Phi(\beta\ell_j)}{\overline{\ell}^2}\frac{\Lambda^2}{(1+\Lambda)^2}\sum_{i,k=1}^N\Phi(\beta\ell_i)\Phi(\beta\ell_k)\ell_i\overline{\mathcal G}_{ik} \ell_k+O(\epsilon^2,r).
\end{align}
We can now safely take the limit $r\rightarrow 0$ for $\epsilon >0$ with $\Lambda \rightarrow \infty$.

\medskip
\noindent (iv) For simplicity, we have considered spherically-shaped targets. However, as originally shown by Ward and Keller \cite{Ward93}, it is possible to generalize the asymptotic analysis of narrow capture problems to more general target shapes such as ellipsoids by applying classical results from electrostatics. In the case of totally absorbing targets one simply replaces the target length $\ell_j$ in the far-field behavior of the inner solution by the capacitance $C_j$ of an equivalent charged conductor with the shape $\calU_j$. In addition, using the divergence theorem, it can then be shown that the flux into a target is completely determined by the far-field behavior. However, the analysis is more complicated when considering partially absorbing targets, since it is necessary to match the solution $U(\y)$ exterior to a target with the solution $V(\y)$ inside the target. The latter would require solving the modified Helmholtz equation in a non-spheroidal shape $\calU_j$.

\subsection{Pair of targets in a sphere}

\begin{figure}[b!]
\begin{center} 
\includegraphics[width=13cm]{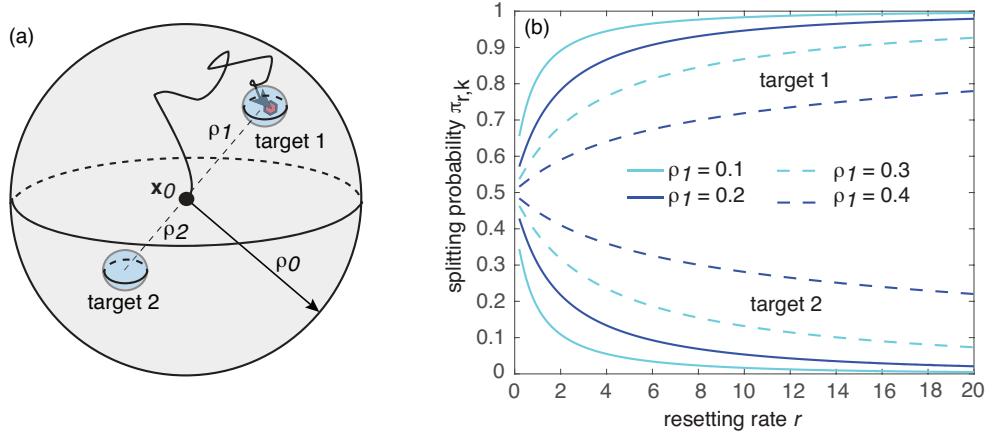} 
\caption{(a) Spherical search domain of radius $r=\rho_0$ containing two diagonally opposed targets at distances $r_j$ from the center. The initial position of the searcher is taken to be at the center of the sphere, $\x_0=0$. (b) Plot of splitting probabilities $\pi_{r,k}(\x_0)$, $k=1,2$, for $\x_0=0$. The splitting probabilities are determined by equations (\ref{Piee}) and (\ref{JLTr}). The distance of the targets are varied such that $\rho_1+\rho_2=1$. Other parameter values are $D=1$, $\ell_j=1$, $\rho_0=1$ and $\epsilon =0.01$. The plots are insensitive to the value of $\kappa_1$ for $\epsilon \ll 1$}
\label{fig5}
\end{center}
\end{figure}

\begin{figure}[b!]
\begin{center} 
\includegraphics[width=13cm]{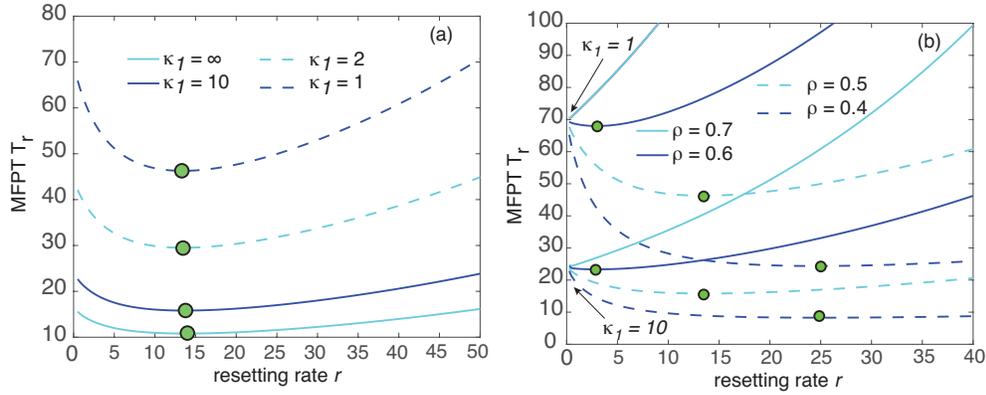} 
\caption{Plot of unconditional MFPT $T_{r}(\x_0)$ as a function of $r$ for the pair of targets shown in Fig. \ref{fig5} with $\x_0=0$. The MFPT is determined by equations (\ref{Ttot}) and (\ref{JLTr}). (a) The distance of the targets are fixed at $\rho_1=\rho_2=0.5$ and the absorption rate $\kappa_1$ is varied.The filled circles indicate the optimal resetting rate for a given $\kappa_1$. (b) Fxed $\kappa_1$ and different target distances $\rho=\rho_{1,2}$. The filled circles indicate the optimal resetting rate for a given $\rho$ and $\kappa_1$. Other parameter values are $D=1$, $\ell_j=1$, $\rho_0=1$ and $\epsilon =0.01$. }
\label{fig6}
\end{center}
\end{figure}

We now combine our asymptotic expansion of the flux given by equation (\ref{JLTr})  with the expressions (\ref{Piee}) and (\ref{Ttot}) for the splitting probability $\pi_r(\x_0)$ and unconditional MFPT $T_r(\x_0)$, respectively, by identifying the Laplace variable $s$ with the resetting rate $r$. If we only include terms to $O(\epsilon^2)$ then
For the sake of illustration, suppose that the search domain is a sphere of radius $\rho_0$. Consider two targets of equal size $\ell_j=1$ located along a diagonal at a distance $\rho_j$ from the center, $j=1,2$, see Fig. \ref{fig5}(a). The Neumann Green's function of the modified Helmholtz equation and of the diffusion equation can be calculated explicitly for a spherical domain as detailed in \cite{Grebenkov20,Bressloff21B}. In Fig. \ref{fig5}(b) we plot the splitting probabilities ${\pi}_{r,k}$, $k=1,2$, as a function of the resetting rate $r$ for $\epsilon \ll r \ll 1/\epsilon$. (The lower bound is imposed due to the singular nature of the Green's functions in the limit $r\rightarrow 0$, see remark (iii).) We also take $\rho_1+\rho_2=1$. Clearly if $\rho_1=\rho_2$ then each the particle is equally likely to be absorbed by either target and $ {\pi}_{r,1}= {\pi}_{r,2}=1/2$. However, if $\rho_1<\rho_2$ then the particle is more likely to be absorbed by the target closer to the origin, that is, $ {\pi}_{r,1}> {\pi}_{r,2}$. We see from Fig. \ref{fig5}(b) that this difference increases with $r$. Analogous results were obtained in our previous study \cite{Bressloff21B}. The additional observation here is that the splitting probabilities are only weakly dependent on the absorption rate $\kappa_1$ and the target size $\epsilon$. Therefore, in order to investigate the effects of partially absorbing targets, we consider the MFPT $T_{r}(\x_0)$. Plots of $T_r(\x_0)$ versus $r$ are shown in Fig. \ref{fig6}(a). The results are similar to the 2D case. For target distances smaller than a critical value $\rho_c$, the MFPT is a unimodal function of $r$ with a minimum at an optimal resetting rate $r_{\rm opt}$. In addition, $r_{\rm opt}$ is only weakly dependent on the absorption rate $\kappa_1$ compared to other parameter values. For example, in Fig. \ref{fig6}(b) we plot $T_r(\x_0)$ against $r$ for different distances $\rho=\rho_{1,2}$, which clearly show a large variation of $r_{\rm opt}$.

\section{Discussion}
In this paper we analyzed the narrow capture problem in 2D and 3D under the joint effects of stochastic resetting and partially absorbing targets. By matching inner and outer solutions of the diffusion equation in Laplace space, we derived an asymptotic expansion of the flux $\widetilde{J}_k$ into each target in powers of $\nu=-1/\ln\epsilon $ in 2D and powers of $\epsilon$ in 3D, where $\epsilon$ is the non-dimensionalized target size. (In the former case, all logarithmic terms were summed over non-perturbatively.) This then determined the MFPT to absorption according to the formula $T_r(\x_0)=(1-  \sum_{k= 1}^N \widetilde{J}_k(\x_0,r))/( r\sum_{j=1}^N\widetilde{J}_j(\x_0,r))$ for a resetting rate $r$. We illustrated the theory by considering spherically symmetric targets. We determined the MFPT to absorption as a function of $r$, $\kappa$ and the target distances from the resetting point, and explored the behavior of the optimal resetting rate as a function of model parameters. 

The analysis of partially absorbing targets developed in this paper has a wide variety of possible extensions. First, as indicated at the end of \S 4.3, one could consider more general target shapes by numerically solving the modified Helmholtz equation within each target and matching with the inner solution outside the target. Second, one could incorporate a more general chemical kinetic scheme within each target, rather than assuming direct absorption. For example, on entering a target, the particle could reversibly bind to the chemical substrate and undergo a sequence of reversible reactions before being absorbed \cite{Schumm21}. Alternatively, the reaction scheme could be non-Markovian, analogous to a study of anomalous diffusion within spiny dendrites \cite{Fedotov08}. In this specific application particles diffuse along a one-dimensional dendritic cable that is studded with partially absorbing spines. The exchange of a particle between a spine and the parent dendrite is described by a non-Markovian stochastic process and leads to subdiffusive transport. Finally, rather than assuming that the boundary of a target is fully permeable to a diffusing particle, one could consider a semi-permeable membrane. The boundary conditions (\ref{master}c) would be replaced by the following equations on $\x \in \partial \calU_i$:
\begin{equation}
 D\nabla q(\x,t|\x_0)\cdot \n_i=-\sigma[q(\x,t|\x_0)-p_i(\x,t|\x_0)] ,\quad D\nabla q(\x,t|\x_0)= D \nabla p_i(\x,t|\x_0)  ,
\end{equation}
where $\sigma$ is a membrane permeability coefficient. A special case of the above is to treat each target as a spatially homogeneous compartment, as exemplified by a model of bacterial quorum sensing \cite{Gou16}.

 \begin{appendix}

 \section{Calculation of $U_2$} In order to calculate the inner contribution $U_2$, we have to match the far-field behavior of $U_2$ with the $O(\epsilon^2)$ term in the expansion of $\q_0$, see equation (\ref{p0}), the $O(\epsilon)$ terms in the expansion of $\q_1$, and the near field behavior of $\q_2 $ around $\calU_j$. The latter satisfies equation (\ref{asym1}) supplemented by the singularity condition
\[\q_2(\x,s)\sim - \Phi(\beta\ell_j)\frac{\chi_j^{(1)}\ell_j}{|\x-\x_j|}, \quad \mbox{as } \x\rightarrow \x_j.\]
Following the same steps as in the derivation of $\q_1(\x,s)$ yields
\begin{equation}
 \q_2(\x,s)= -{4\pi}D\sum_{k=1}^N \Phi(\beta\ell_k)\chi_k^{(1)}\ell_kG(\x,s|\x_k) .
\end{equation}
Hence,
\begin{align}
U_2(\y,s)\rightarrow  \frac{1}{2}{\bf H}_j\cdot \y\otimes \y+\left . \nabla \tilde{q}_1 \right |_{\x=\x_j}\cdot \y-4\pi D\Phi(\beta\ell_j)\sum_{k=1}^N\chi_k^{(1)}\ell_k{\mathcal G}_{jk} 
\end{align}
as $ |\y|\rightarrow \infty $. In addition, setting $n=2$ in equation (\ref{stretch}) we have
\begin{align}
D\Delta_{\y} U_2 =sU_0,\  |\y|>\ell_j,\quad D\Delta_{\y} V_2 -\kappa_1 V_2=sV_0,\  |\y|<\ell_j,\\
U_2=V_2,\quad \nabla U_2\cdot \n_j =\nabla V_2 \cdot \n_j \mbox{ on } |\y|=\ell_j. 
\end{align}
 Again we decompose the inner term as
$U_2=W_2+\widehat{W}_2$ with 
\begin{align}
\label{chi2}
\Delta_{\y}W_2&=0, |\y|>\ell_j; \    W_2\rightarrow \chi_j^{(2)}\equiv -4\pi D\Phi(\beta\ell_j)\sum_{k=1}^N\chi_k^{(1)}\ell_k{\mathcal G}_{jk}\mbox{ as } |\y|\rightarrow \infty,\end{align}
and
\begin{align}
\label{Bj}
D\Delta_{\y}\widehat{W}_2&=sU_0, \ |\y|>\ell_j;\  \widehat{W}_2=0 ,\mbox{ on } |\y|=\ell_j,\\ \widehat{W}_2&\rightarrow \frac{1}{2}{\bf H}_j\cdot \y\otimes \y +\left . \nabla \tilde{q}_1 \right |_{\x=\x_j}\cdot \y\mbox{ as } |\y|\rightarrow \infty.\nonumber
\end{align}
The general solutions for $W_2$ and $V_2$ are
\begin{equation}
W_2= \chi_j^{(2)}+{A_2}\frac{\ell_j}{|\y|},\quad V_2=B_2\frac{I_{1/2}(\beta|\y|)}{\sqrt{\beta|\y|}}+s\widehat{V}_2(\y,s;\kappa_1),
\end{equation}
where $\widehat{V}_2$ is the particular solution with boundary condition $\widehat{V}_2(\y,s;\kappa_1)= 0$ on $|\y|=\ell_j$ (convolution of the Green's function in the sphere with $V_0$.) The explicit expression for $\widehat{V}_2(\y,s;\kappa_1)$ will not be needed in this paper.

The calculation of $\widehat{W}_2$ proceeds along analogous lines to \cite{Bressloff21B}, so we simply summarize the results here. First note that the particular solution of the equation $D\Delta_{\bf y}\widehat{W}_2=sU_0$ has the explicit form
\begin{equation}
\label{B2}
\left .\widehat{W}_2\right |_{\rm particular} =\frac{s}{D}\left (\frac{\rho^2}{6}-\frac{\rho\ell_j\Phi(\beta\ell_j)}{2}\right )G(\x_j,s|\x_0).
\end{equation}
 It turns out that this matches the $l=0$ component of the far-field condition, since
\begin{align*}
 \frac{1}{2}{\bf H}_j\cdot \y\otimes \y+\nabla \q_1\cdot \y =\frac{\rho^2}{6}(H_{xx}+H_{yy}+H_{zz})+\mbox{ higher-order spherical harmonics},
\end{align*}
and
\[\frac{\rho^2}{6}(H_{xx}+H_{yy}+H_{zz})=\frac{\rho^2}{6}\Delta G(\x_i,s|\x_0)=\frac{\rho^2s}{6D}G(\x_i,s|\x_0).\]
Hence, the only contribution to the flux integral from the component $\widehat{W}_2$ arises from the above particular solution.
Finally, the unknown constants $A_2$, $B_2$ are determined by the boundary conditions at $|\y|=\ell_j$:
\begin{align}
\chi_j^{(2)}+{A_2}&=B_2I_{1/2}(\beta\ell_j)/\sqrt{\beta\ell_j},\\
 -\frac{A_2}{\ell_j} +\frac{s\ell_j}{6D}\left (2-3  \Phi(\beta\ell_j) )\right )G(\x_j,s|\x_0)+\Theta_{2,j}&=\frac{B_2}{\sqrt{\beta \ell_j}}\left [\beta I'_{1/2}(\beta\ell_j)  -\frac{1}{2\ell_j} I_{1/2}(\beta\ell_j) \right ]\nonumber \\
 &\qquad+\widehat{V}_2'(\ell_j,s;\kappa_1).\nonumber
\end{align}
We have defined
\begin{equation}
\Theta_{2,j}=\n_j\cdot \nabla \left (\widehat{W}_2-\left .\widehat{W}_2\right |_{\rm particular}\right )_{|\y|=\ell_j},
\end{equation}
which collects together all terms involving higher-order spherical harmonics.

\end{appendix}

\end{document}